\documentclass[10pt,conference, compsocconf]{IEEEtran}

\usepackage[svgnames, x11names]{xcolor}

\usepackage[normalem]{ulem} 

\usepackage{listings}

\lstdefinelanguage{XML}
{
basicstyle=\ttfamily\footnotesize,
  morestring=[b]",
  moredelim=[s][\bfseries\color{Maroon}]{<}{\ },
  moredelim=[s][\bfseries\color{Maroon}]{</}{>},
  moredelim=[l][\bfseries\color{Maroon}]{/>},
  moredelim=[l][\bfseries\color{Maroon}]{>},
  morecomment=[s]{<?}{?>},
  morecomment=[s]{<!--}{-->},
  commentstyle=\color{gray},
  stringstyle=\color{blue},
  identifierstyle=\color{red}
}
\usepackage[pdftex]{graphicx}
\graphicspath{{./figures/}}
\DeclareGraphicsExtensions{.pdf}

\usepackage[cmex10]{amsmath}
\usepackage{amssymb}
\usepackage{mathtools}

\usepackage{subfig}

\usepackage{algorithmicx}
\usepackage{algpseudocode}
\usepackage[ruled]{algorithm}
\definecolor{light-gray}{gray}{0.75}
\algrenewcommand{\algorithmiccomment}[1]{\hskip3em{{\footnotesize \textcolor{light-gray}{$\blacktriangleright$}}} #1}

\usepackage{multirow}

\usepackage{hyperref}

\usepackage{xspace}
\usepackage[nocompress]{cite}

\usepackage{enumitem}

\IEEEoverridecommandlockouts
\begin{document}
%
\title{Distributed Algorithms for Subgraph-Centric Graph Platforms}
%
%
%
%

\author{\IEEEauthorblockN{Diptanshu Kakwani$^{1,*}$\thanks{$^{*}$Based on work done as an intern at the DREAM:Lab, CDS, IISc} and Yogesh Simmhan$^2$}
\IEEEauthorblockA{$^1$Information Science and Engineering, M. S. Ramaiah Institute of Technology, Bangalore\\
$^2$Department of Computational and Data Sciences, Indian Institute of Science, Bangalore\\
Email: dipkakwani@gmail.com, simmhan@cds.iisc.ac.in}
}

\maketitle

\begin{abstract}
Graph analytics for large scale graphs has gained interest in recent years. Many graph algorithms have been designed for vertex-centric distributed graph processing frameworks to operate on large graphs with $100~M$ vertices and edges, using commodity clusters and Clouds. Subgraph-centric programming models have shown additional performance benefits than vertex-centric models. But direct mapping of vertex-centric and shared-memory algorithms to subgraph-centric frameworks are either not possible, or lead to inefficient algorithms. In this paper, we present three subgraph-centric distributed graph algorithms for triangle counting, clustering and minimum spanning forest, using variations of shared and vertex-centric models. These augment existing subgraph-centric algorithms that exist in literature, and allow a broader evaluation of these three classes of graph processing algorithms and platforms. 
\end{abstract}


\section{Introduction}
With the ever-growing size of data, the problem of processing large datasets that do not fit in a single machine has become important. Graphs constitutes a particularly challenging class of data due to their irregular structure and asymmetric execution of graph algorithms. Distributed Big Data platforms like MapReduce are ill-suited for graph computation~\cite{cohen:aip:2009} while parallel libraries like BOOST~\cite{lumsdaine:boost:2002} rely on costly HPC clusters. Google introduced a vertex-centric distributed graph processing model called Pregel~\cite{malewicz:sigmod:2010} for processing large scale graphs on commodity clusters. Pregel uses a Bulk Synchronous Parallel (BSP) iterative programming model and its application logic is written from the perspective of individual vertices. Apache Giraph is an open source implementation of Pregel. Other frameworks have extend this to a coarser component-centric programming model, including GoFFish (subgraph-centric)~\cite{simmhan:europar:2014}, Giraph++ (partition-centric)~\cite{tian:vldb:2013} and Blogel (block-centric)\cite{yan:vldb:2014}. These have exhibited better performance than Pregel by reducing the number of supersteps required for the application to complete and the number of messages exchanged between workers. 

However, the body of subgraph-centric algorithms is limited, and it is not possible to na\"{i}evely map from vertex-centric and shared-memory graph algorithms to their component-centric counterparts. In our previous work on GoFFish~\cite{simmhan:europar:2014}, we presented subgraph-centric algorithms for connected-components, single-source shortest path, PageRank, Subgraph-Rank~\cite{badam:sigmod:2014} and Approximate Betweenness Centrality~\cite{dindokar:ccgrid:2016}. Others have performed subgraph-centric partitioning and have presented algorithms for connected-components~\cite{tian:vldb:2013}, graph coarsening~\cite{tian:vldb:2013}, PageRank~\cite{tian:vldb:2013,yan:vldb:2014}, single-source shortest path\cite{yan:vldb:2014} and reachability~\cite{yan:vldb:2014} as well. In this paper, we augment the existing suite of subgraph-centric algorithms with three new distributed ones for triangle-counting, k-way clustering and minimum spanning forest, while drawing upon their shared-memory and vertex-centric formulations. We also empirically validate the triangle counting algorithm, and confirm its superior performance to the vertex centric equivalent.

In the rest of the paper, we briefly review related work~\ref{sec:related}, present the triangle counting~\ref{sec:triangle}, k-way clustering~\ref{sec:kways} and minimum spanning forest~\ref{sec:msf} algorithms, and discuss experimental results for triangle counting for 3 real-world graphs~\ref{sec:results}.

\section{Background and Related Work}
\label{sec:related}
In this section, we discuss related work on distributed graph processing frameworks and algorithms, and provide background on the subgraph-centric programming abstractions that we leverage from the GoFFish framework for our proposed algorithms.

Apache Giraph is an open source implementation of Pregel~\cite{malewicz:sigmod:2010}, which is a vertex-centric distributed graph processing framework. Pregel and other component-centric distributed graph processing frameworks like Giraph++, Blogel and GoFFish~\cite{tian:vldb:2013,simmhan:europar:2014,yan:vldb:2014}, uses a Bulk Synchronous Parallel (BSP)~\cite{valiant:acm:1990} execution model. BSP is an iterative computation model where a superstep (iteration) consists of a user-specified function (\texttt{compute()}) run in parallel by workers on different parts of the graph, with the functions optionally generating messages to other parts of the graph. This is followed by bulk message transfer between the workers with a barrier synchronization at the end. As the supersteps repeat, the user's compute function has access to messages published to it from other workers in the previous superstep. These supersteps repeat until the application completes, which is identified by a global consensus when all workers have \emph{voted to halt}, and no messages are in-flight. 

In a vertex-centric programming model, the user's \texttt{compute()} function is run on every vertex of the partitioned graph whereas in a subgraph-centric model, it is run on every subgraph (weakly connected component) present in each partition of the graph. Messages can likewise be targeted from a vertex/subgraph to a neighboring remote vertex/subgraph. For e.g., Table~\ref{tbl:methodspecs} illustrates the programming abstractions provided by the GoFFish platform which we use to define the subgraph-centric algorithms in the upcoming sections.
\begin{table}[h]
	\centering
	\caption{Programming Abstractions Provided by GoFFish} 
	\begin{tabular}{l||p{5.8cm}}
		\hline
		Function & Description \\
		\hline
		\hline
		\texttt{{Compute}} &  User specified function to be run on each subgraph, for each superstep \\
		\texttt{{Send}} &  Sends a message to a remote subgraph, or vertex in a subgraph \\
		\texttt{{SendToAll}} &  Broadcasts a message to all subgraphs in the graph \\
		\texttt{{SendToMaster}} &  Sends message to the \emph{master} subgraph \\
		\texttt{{VoteToHalt}} & Subgraph votes to halt. Application halts when all subgraphs halt and no messages were sent.  \\
		\hline
	\end{tabular}
	\label{tbl:methodspecs}
\end{table}

Many algorithms have been designed and implemented for both vertex-centric frameworks~\cite{malewicz:sigmod:2010,salihoglu:ssdbm:2013,salihoglu:vldb:2014} and subgraph-centric frameworks~\cite{tian:vldb:2013,yan:vldb:2014,badam:sigmod:2014}. \cite{malewicz:sigmod:2010} proposed algorithms for PageRank, single source shortest path, bipartite matching and semi-clustering for Pregel. \cite{salihoglu:vldb:2014} presented optimized algorithms for strongly connected components, minimum spanning forest, graph coloring, approximate maximum weight matching and weakly connected components for Pregel-like systems. Blogel~\cite{yan:vldb:2014} offered block-centric algorithms for single source shortest path, reachability and PageRank. Giraph++~\cite{tian:vldb:2013} developed graph-centric algorithms for connected components, PageRank and graph coarsening. Our prior work on GoFFish~\cite{simmhan:europar:2014} included subgraph-centric algorithms for connected components, single source shortest path, PageRank and Subgraph-Rank\cite{badam:sigmod:2014}. Often, the component-centric algorithms are inspired both by shared-memory algorithms, as well as their vertex-centric variations. For e.g., the SSSP algorithm in GoFFish switches between Dijkstra's and vertex-centric in each superstep to achieve a subgraph-centric formulation.

While these earlier works have demonstrated the performance improvement of component-centric graph algorithms over vertex centric algorithms, there are still a large swathe of algorithms that do not yet have a subgraph-centric version that would benefit from the performance and scalability benefits. We address this gap here. To the best of our knowledge, the algorithms which we present in this paper have not been designed for any component-centric abstraction previously.

\section{Triangle Counting}
\label{sec:triangle}

A \emph{triangle} in a graph is a 3-clique, that is, a set of three vertices where each pair of vertices is connected. \emph{Triangle counting} is the problem of finding all triangles in a graph. The application of triangle counting includes finding clustering coefficient, transitivity, and community detection~\cite{kolountzakis:waw:2010}. This problem has been studied extensively for shared-memory system. \cite{schank:iweea:2005} survey existing shared-memory algorithms, with runtime complexities of $\mathcal{O}(m^\frac{3}{2})$ and $\mathcal{O}(d_{max}^2.n)$, where $n, m$ and $d_{max}$ are respectively the number of vertices, edges, and maximum degree in the graph.

A vertex-centric algorithm for triangle counting has been proposed by~\cite{ediger:ipdpsw:2013}, and is performed in three supersteps: (i) Each vertex sends its ID as a message to its neighboring vertices having a higher ID than itself, (ii) A vertex that receives a message in the second superstep from the first forwards the message to neighbors with a higher id than itself, and (iii) If the vertex ID received in the message from superstep 2 in superstep 3 is present in the adjacency list of the receiving vertex, then a triangle is formed. This algorithm is simple but suffers from high communication costs due to $\mathcal{O}(m)$ messages that are passed after each of the first two supersteps in this vertex-centric formulation.

\begin{figure}[h]
\centering
	\subfloat[Type(i)\label{subfig-1:type1}]{%
		\includegraphics[width=0.21\columnwidth]{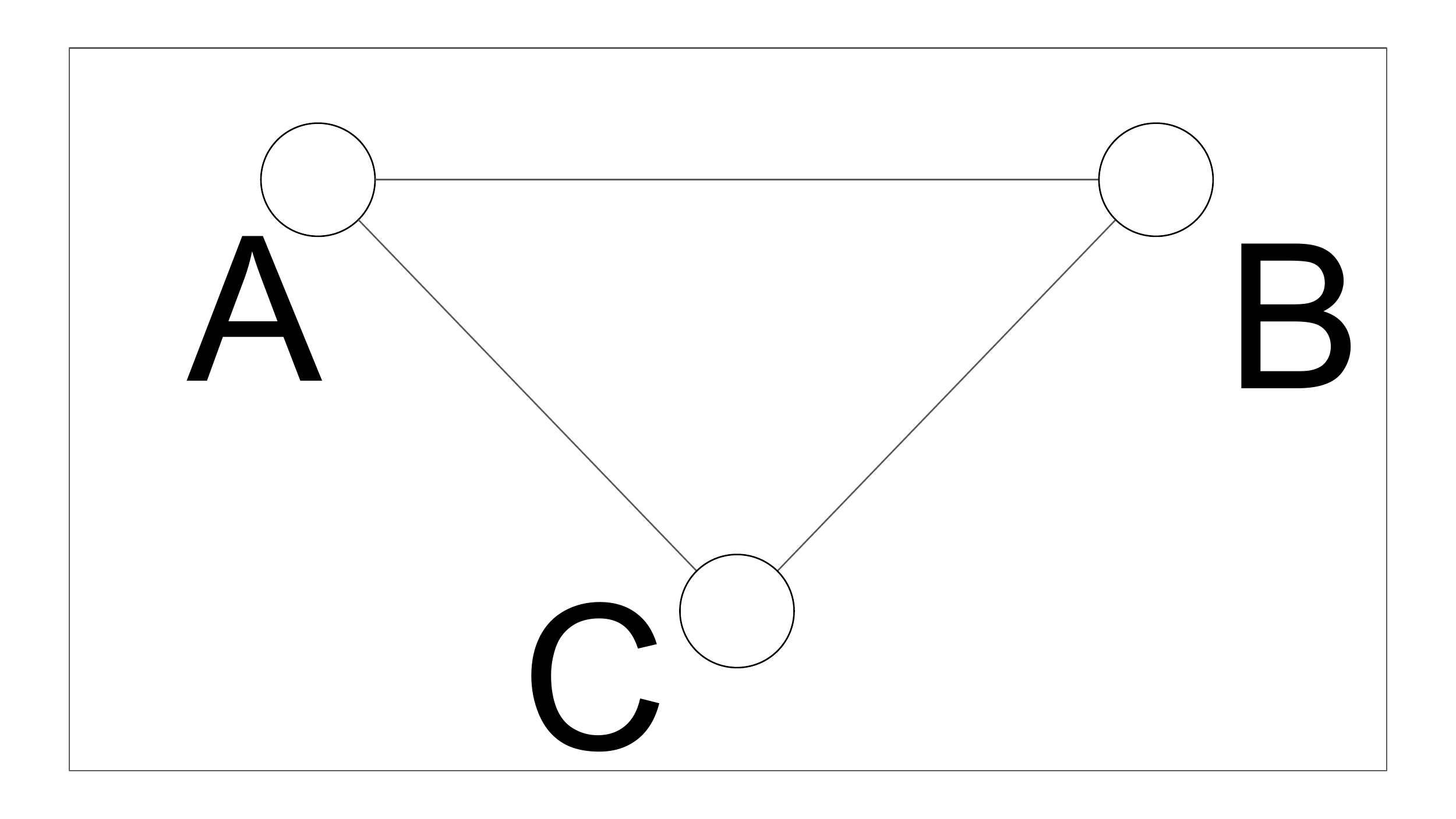}
	}~~
	\subfloat[Type(ii)\label{subfig-2:type2}]{%
		\includegraphics[width=0.28\columnwidth]{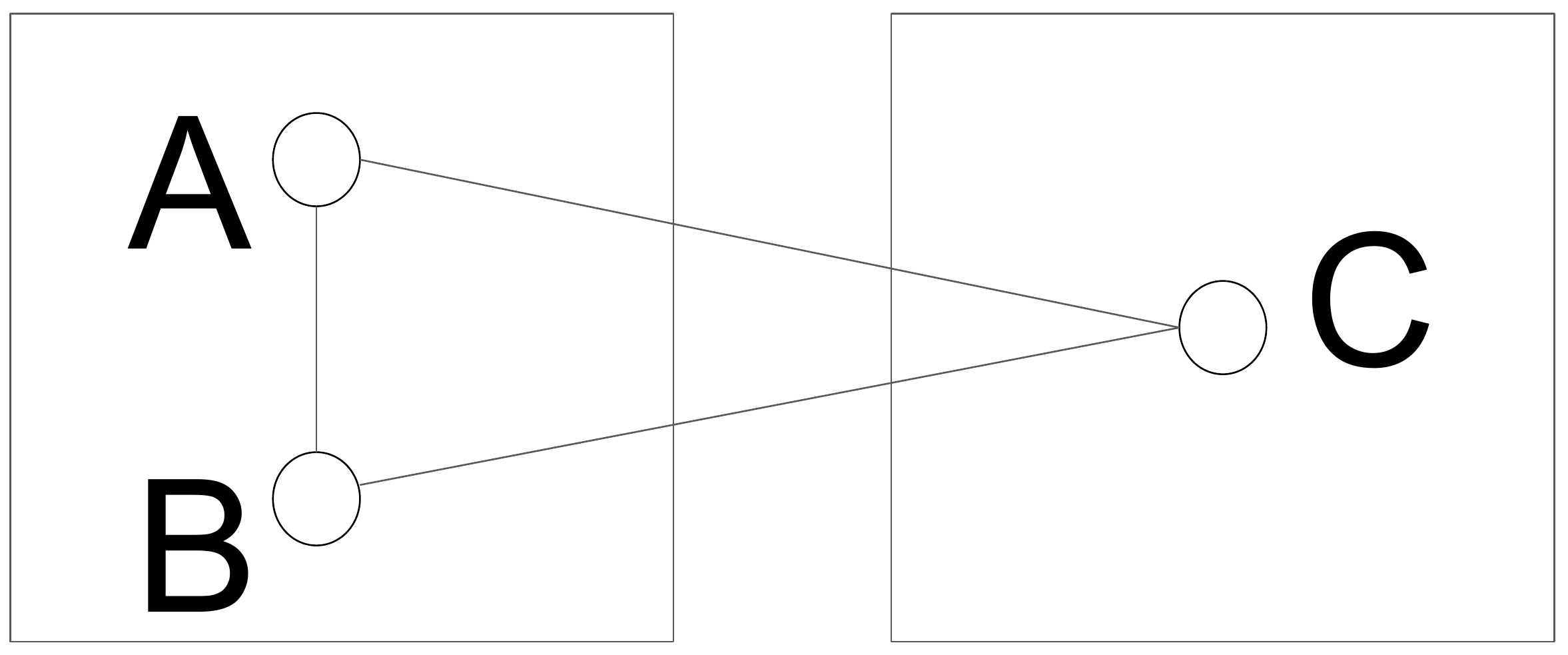}
	}~~
	\subfloat[Type(iii)\label{subfig-3:type3}]{%
		\includegraphics[width=0.30\columnwidth]{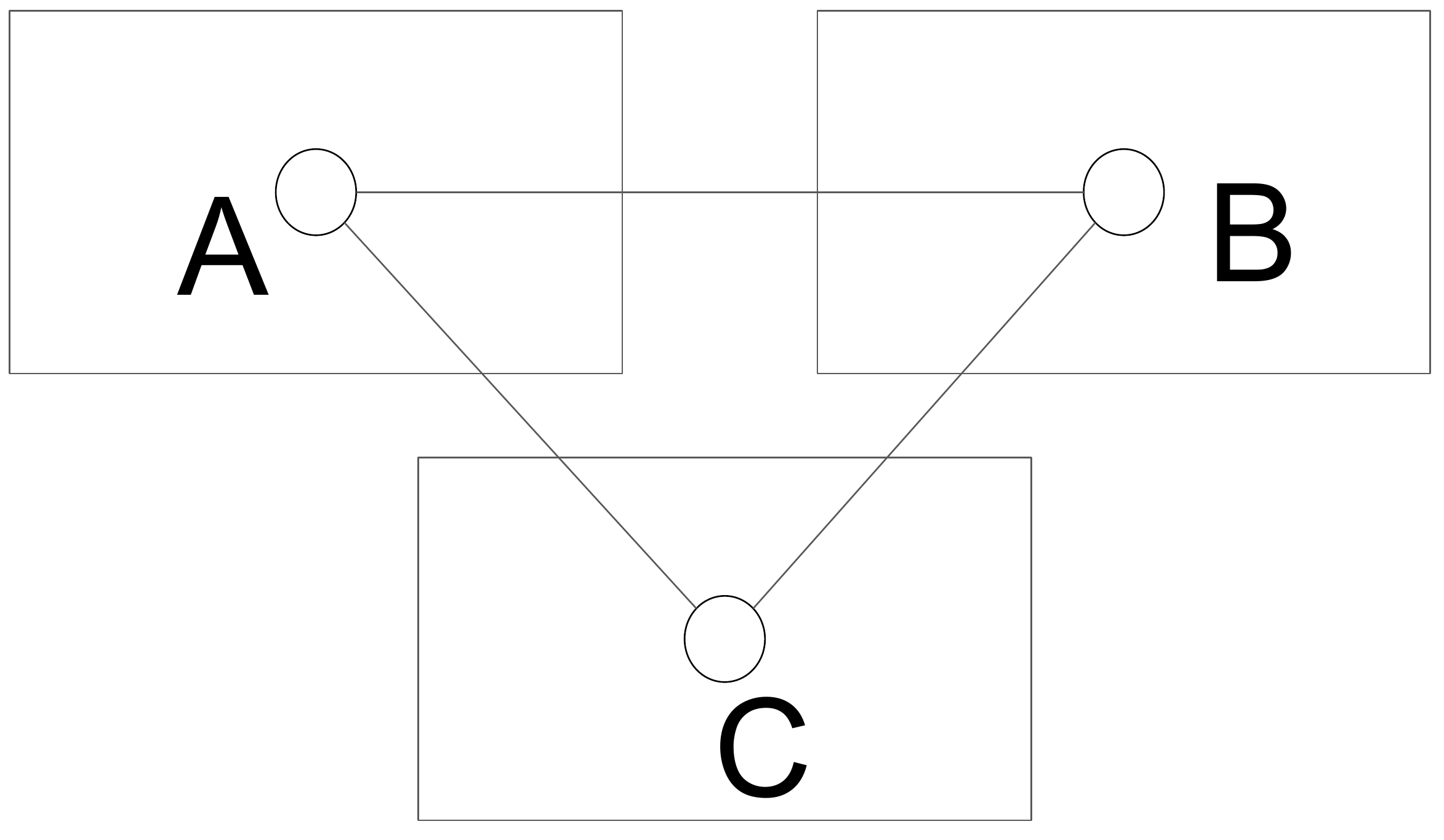}
	}
	\caption{Types of triangle spread across workers.}
	\label{fig:triangle}
\end{figure}

The subgraph-centric algorithm which we present avoids this high communication complexity by making use of the information available within the subgraph in each superstep. It limits communication to the boundary vertices that may participate in triangles that span subgraphs. Any triangle can be classified into three types based on the location of its vertices with respect to the subgraphs: (i) all three vertices lie in the same subgraph, (ii) two of its vertices lie in one subgraph, while the third is in a different subgraph, and (iii) all three vertices lie in different subgraphs. Types (i) and (ii) can be identified using the information available within one subgraph in just one superstep. For type (i), this is similar to a shared memory algorithm, and even for type (ii), the two boundary vertices from the first subgraph have access to the vertex ID of their neighboring vertex (though not other properties) in a remote subgraph, allowing them to test for a shared neighbor that completes a triangle. However, for type (iii) we need three supersteps, as for that of a vertex-centric algorithm. These types are illustrated in Fig.~\ref{fig:triangle}.

\begin{algorithm}[t]
	\footnotesize
	\caption{Triangle Counting}
	\label{alg:triangle-count}
	\begin{algorithmic}[1]
		\Procedure{Compute}{Subgraph s, Message[] msgs} 
		\If{superstep = 0} \textit{\Comment Find Types (i),(ii) triangles in subgraph}\label{superstep0start}
			\For{$v \in s.vertices$} \label{superstep0type12start}
				\For{$w \in  v.adjList~|~!w.isRemote \text{ \& } w.id > v.id$ }
					\For{$u \in w.adjList~|~u.id > w.id$}
						\If{$u \in v.adjList$}\emph{\Comment v, w, u form triangle}
							\State \textsc{Print}$(v.id, w.id, u.id)$
						\EndIf
					\EndFor
				\EndFor
			\EndFor \label{superstep0type12end}
			\For{$v \in s.vertices$} \label{superstep0type3start}
				\For{$w \in  v.adjList~|~w.isRemote \text{ \& } w.id > v.id$ }
\textit{\Comment Potential Type (iii). Message subgraph 2.}
                                \State $\textsc{Send}(w.subgraphId, \langle v.id, w.id \rangle)$
                                \EndFor
			\EndFor \label{superstep0end}
		\ElsIf{superstep = 1}  \label{superstep1start}
			\For{$\langle vId, wId \rangle \in msgs$}
				\State{$w = s.getVertex(wId)$ }
				\For{$u \in  w.adjList~|~u.isRemote \text{ \& } u.id > w.id$ }\label{edge-existence}
\textit{\Comment Potential Type (iii). Message subgraph 3.}
                                \State $\textsc{Send}(u.subgraphId, \langle vId, wId, u.id \rangle)$
                                \EndFor
			\EndFor \label{superstep1end}
		\ElsIf{superstep = 2} \label{superstep2start}
			\For{$\langle vId, wId, uId \rangle \in msgs$}
				\State $u = s.getVertex(uId)$
				\If{$vId \in u.adjList$} \emph{\Comment Found Type (iii) triangle}
					\State \textsc{Print}$(vId, wId, uId)$
				\EndIf
			\EndFor
		\EndIf \label{superstep2end}
		\EndProcedure
	\end{algorithmic}
\end{algorithm}

Algorithm~\ref{alg:triangle-count} shows our proposed subgraph-centric triangle counting algorithm. The algorithm assumes the graph is undirected and each vertex has an adjacency list of its neighboring vertices ($adjList$) that can be looked up in constant time. In the first superstep, it begins with finding and printing triangles whose vertices ($v,w,u$) are fully internal to this subgraph, or with just one vertex ($u$) in a remote subgraph (lines~\ref{superstep0type12start}--\ref{superstep0type12end}). It then sends messages to vertices in neighboring subgraphs whose IDs ($w$) are larger than the source vertex ($v$) in this subgraph (lines~\ref{superstep0type3start}--\ref{superstep0end}).  In the second superstep (lines~\ref{superstep1start}--\ref{superstep1end}), the received message with a pair of vertex IDs $\langle v,w \rangle$ is forwarded to a neighboring subgraph connected to the local vertex ($w$) if the remote subgraph's vertex ID ($u$) is higher than the source vertex ($w$). In the last superstep (lines~\ref{superstep2start}--\ref{superstep2end}), we receive vertex IDs for a triple $\langle v,w,u \rangle$ and test if $u$ is connected to $v$, and if so, print it as a Type (iii) triangle. Note that in both supersteps 2 and 3, one may end up not finding a Type (iii) triangle if the vertex ID comparisons fail ($w.id < v.id$, $u.id < w.id$), or the first vertex is not connected to the third ($vId \not\in u.adjList$). 

Say $l_{max}$ is the largest number of local vertices among all subgraphs and $r_{max}$ the largest number of edge cuts in all subgraphs. As can be seen, the computational complexity of this algorithm is $\mathcal{O}(d_{max}^2.l_{max})$ and the communication complexity is $\mathcal{O}(r_{max})$, which are much smaller than the vertex-centric algorithm. With a reasonable partitioning of the graph into $p$ subgraphs, we expect $l_{max} \approx \frac{n}{p}$ and $r_{max} \ll m$, and so this gives us a linear speedup over the shared-memory algorithm as well.

\section{k-Way Clustering}
\label{sec:kways}

\cite{salihoglu:ssdbm:2013} has proposed a randomized vertex-centric algorithm for clustering a graph that behaves like k-means clustering, we refer to \emph{k-way clustering}. The algorithm finds $k$ clusters in the graph such that the vertices in the cluster are connected, no two clusters share a vertex, and the number of edge cuts between clusters is smaller than a given threshold. 

The iterative algorithm works across 4 phases as follows: (i) Pick $k$ random vertices as cluster centers; (ii) assign each vertex in the graph to the nearest cluster center by performing a breadth first search (BFS) from each center; (iii) count the number of edges crossing clusters; (iv) if the edge cuts are greater than the threshold, return to phase (i) to pick $k$ random vertices and repeat the algorithm. Each phase may run across multiple supersteps. We adapt this algorithm to a subgraph-centric model, shown in Algorithm~\ref{alg:k-means}. 

The first phase selects $k$ random vertices as cluster centers using the distributed reservoir sampling algorithm~\cite{vitter:toms:1985}. Here, each subgraph picks $k$ local vertices and assign them each a randomly generated key in the first superstep (lines~\ref{alg2:samplingstart}). These vertex IDs and their keys are then sent to all other subgraphs. In the second superstep, each subgraph independently selects the (same) top $k$ vertices from this list as the cluster centers by sorting them on their random keys (line ~\ref{global_vertices}). 

The next phase assigns each vertex in the graph to one of the $k$ cluster centers. We do this by performing a subgraph-centric multi-source BFS starting concurrently from each of the $k$ centers, and each vertex being labeled with the center vertex it is closest to, assuming unit-weighted edges. The \texttt{localBFS} starts a traversal from vertex centers in that subgraph, and when boundary vertices connecting to other subgraphs are reached, pass messages to the remote vertex with the distance(s) from the source(s). This repeats until no vertex distances and labels change (Lines \ref{local-bfs-start}-\ref{local-bfs-end}). The number of supersteps in this phase is upper-bounded by the diameter of the meta-graph, where subgraphs are meta-vertices and remote edges are meta-edges (which can be much smaller than the diameter of the entire graph)~\cite{simmhan:europar:2014}. 
While the original multi-source BFS halts the program once converged, we instead assign one subgraph as a master which receives notifications from all subgraphs which are not updated and the master decides the phase can change when all vertices in subgraphs have been assigned (lines~\ref{masterstart}--\ref{masterend}).

Once the vertices are assigned to centers and thus form clusters, we calculate the edge cuts between clusters. We first send the center ID for boundary vertices to their remote neighbors having a larger vertex ID than themselves (line \ref{edgecutmsg}). Then, in the next superstep, they count the edge cuts between vertices belonging to different centers, both for local vertices and, using the input messages, with remote vertices (lines~\ref{edgecrossingstart}--\ref{edgecrossingend}). This count is broadcast to all other subgraphs which then  decide if the cut count is below a threshold value, $\tau$, in which case they halt; otherwise, they restart the selection of random $k$ centers. 

\begin{algorithm}[t]
	\footnotesize
	\caption{k-way Clustering}
	\label{alg:k-means}
	\begin{algorithmic}[1]
		\Procedure{compute}{Subgraph s, Messages msgs[]}
			\If{phase = RANDOM\_K\_LOCAL} 
                        \State \textit{\Comment{Select $k$ random local vertices with random keys}}				
				\State localCenters[~] = \textsc{selectRandom}(s.vertices, k)\label{alg2:samplingstart}
				\State \textsc{sendToAll}(localCenters)
				\State phase = TOP\_K\_GLOBAL \textit{\Comment{Change phase}}
			\ElsIf{phase = TOP\_K\_GLOBAL}
                        \State \textit{\Comment{Select top $k$ vertices with smallest keys}}
				\State centers[~] = \textsc{selectTopK}(msgs, k)\label{global_vertices}
                                \State \emph{\Comment{Send BFS start vertices to self, distance 0}}
				\State \textsc{Send}(s.id, $\langle c, 0 \rangle$) $\forall c \in centers$ \label{initial-bfs-vertices}
				\State phase = ASSIGN\_CLUSTER \textit{\Comment{Change phase}}
			\ElsIf{phase = ASSIGN\_CLUSTER}
				\State localUpdates[~] = $\langle v, d \rangle ~\mid~ \langle v, d \rangle \in msgs$ \label{local-bfs-start}
                                \State remoteUpdates[~] = \textsc{localBFS}(localUpdates) 
				\If{remoteUpdates.count $>$ 0}
                                \State \textsc{Send}(sid, $\langle v, d \rangle$) $\forall \langle sid, v, d \rangle \in$ remoteUpdates[~] \label{local-bfs-end}
                                \Else \State \textsc{sendToMaster}(ASSIGNED)\label{masterstart}
				\EndIf
				\If {s.isMaster \& \textsc{Count}(ASSIGNED $\in msgs$) = S\_COUNT}
					\State \textsc{SendToAll}(phase=EDGE\_CUT)
				\EndIf\label{masterend}
			\ElsIf{phase = EDGE\_CUT} \textit{\Comment{Notify remote subgraphs}}
				\State \textsc{Send}($rs.id, \langle v_j, v_i.center \rangle$) $\forall e=(v_i, v_j) \in s ~\&~ v_i \in s ~\&~ v_j \in rs ~\&~ v_j.id > v_i.id$ \label{edgecutmsg}
				\State phase = EDGE\_COUNT
			\ElsIf{phase = EDGE\_COUNT} \label{edgecrossingstart} \textit{\Comment{Sum local, remote cuts}}
                                \State localCuts = \textsc{Count}($v.center \neq w.center ~\forall~ \langle v, w \rangle \in s.edges$)
				\State remoteCuts = \textsc{Count}($v.center \neq rc ~\forall~ \langle v, rc \rangle \in msgs$)
				\State \textsc{SendToAll}(localCuts + remoteCuts)\label{edgecrossingend}
				\State phase = FINISH
			\ElsIf{phase = FINISH}
				\If{\textsc{SumEdgeCuts}(msgs)  $> \tau$}
					\State phase = RANDOM\_K\_LOCAL
				\Else
					\State \textsc{VoteToHalt}() \textit{\Comment{Good quality clusters}}
				\EndIf
			\EndIf
		\EndProcedure
	\end{algorithmic}
\end{algorithm}

\section{Minimum Spanning Forest}
\label{sec:msf}

Minimum spanning tree is a spanning tree of connected, undirected graph which has minimum sum of weights of its edges. Minimum spanning forest (MSF) of an undirected graph is the union of minimum spanning trees for its connected components. We present an algorithm which is based on parallel version of Boruvka's MSF algorithm~\cite{chung:ipps:1996} and its vertex-centric formulation~\cite{salihoglu:vldb:2014}. The algorithm maintains a forest which initially has each vertex as a tree. It then finds minimum outgoing edge from all the trees and merges trees which are connected by those edges. It continues till either one tree is formed or there is no outgoing edge.

Algorithm~\ref{alg:msf} shows the subgraph-centric algorithm for minimum spanning forest. We run the Boruvka's shared-memory algorithm within a subgraph until we hit the boundary vertices of the subgraph. Boruvka's operates by iteratively selecting the smallest weighted edge from pairs of components that are not connected, and growing the component. The outcome is a forest of trees in the subgraph, with the vertex with the smallest ID in a tree serving as the root for vertices in that tree (line \ref{alg:boruvkas:start}-\ref{alg:boruvkas:end}). The root will inherit the adjacency list of its children and serve as the proxy for sending and receiving messages on behalf of them.

Next, we attempt to grow the trees by finding the smallest edges that connect trees in different subgraphs. In one superstep, each root sends a \emph{question} message to its remote edge with the smallest weight (line \ref{alg:msf:question}). In the next superstep, once roots have received messages, they start merging roots that are connected (line \ref{alg:msf:pjump}). They check if they have mutually exchanged questions with another root. If so, the two trees are conceptually merged and the root with the smaller vertex ID becomes the new root for this merged tree. 

If a root did not receive a question from another root it had sent a question to, it merges with the other root as its parent. This merge can have a cascading effect when a root merges into a second root, in the same superstep as this second root merges into a third root, and so on. Using ``pointer jumping''~\cite{salihoglu:vldb:2014}, this converges in $\mathcal{O}(log(d))$ supersteps, where $d$ is the longest distance from the eventual root to all other root vertices that are merging with it to form a single tree in this phase. At the end of this phase, which is detected using a master subgraph, we have a smaller number of larger trees, with a set of corresponding roots that have inherited the adjacency list of their children (lines \ref{alg:msf:merge:start}-\ref{alg:msf:merge:end}).

This above phase then repeats, with the new roots sending a \emph{question} message on the edge with the least weight, and so on till no remote edge exists to another root, if any (line \ref{alg:msf:next}).

\algdef{SE}[DOWHILE]{Do}{doWhile}{\algorithmicdo}[1]{\algorithmicwhile\ #1}%
\begin{algorithm}[t]
	\footnotesize
	\caption{Minimum Spanning Forest (MSF)}
	\label{alg:msf}
	\begin{algorithmic}[1]
		\Procedure{compute}{Messages msgs[~]}
		\If{phase = LOCAL\_MSF} 
                  \State roots[~] = s.vertices[~]
                  \Do \textit{\Comment{Do local MSF using Bovruka's}}\label{alg:boruvkas:start}
                    \State roots[~] = \textsc{BoruvkasMSF}(roots) 
                  \doWhile{$\exists \textsc{MinEdge}(root).isLocal ~|~ root \in roots[~] $} \label{alg:boruvkas:end}
                  \State phase = QUESTION\_REMOTE \textit{\Comment{Change phase}}
                \ElsIf{phase = QUESTION\_REMOTE}

\Comment \emph{Inform remote edge with least weight from each root}
                  \State \textsc{SendMessage}$(e.subgraphId, \langle e.sinkId, r.id \rangle)$ $|~ e=\textsc{MinEdge}(r) ~\&~ e.isRemote ~\forall~ r \in roots[~]$ \label{alg:msf:question}
                  \State phase = MERGE\_ROOTS \textit{\Comment{Change phase}}
		\ElsIf{phase = MERGE\_ROOTS}

\Comment \emph{Pointer jumping to merge spanned roots and find new root}
                  \State done = \textsc{MergeRoots}() \label{alg:msf:pjump}
                  \If{done}  \textit{\Comment{Done local merges}} \label{alg:msf:merge:start}
                    \State \textsc{SendToMaster}(MERGE\_DONE)
                    \State \textsc{SendAdjList}()\textit{\Comment{Send adjacency list to merged root}}
                  \EndIf
                  \If {s.isMaster \& \textsc{AllDone}(MERGE\_DONE)}

\textit{\Comment{Done global merges}}
                    \State \textsc{SendToAll}(phase=NEXT\_ITER)
                  \EndIf \label{alg:msf:merge:end}
		\ElsIf{phase = NEXT\_ITER} 

\Comment \emph{More remote edges exist? Repeat.}
                  \If{$\exists \textsc{MinEdge}(root).isRemote ~|~ root \in roots[~] $} \label{alg:msf:next}
                    \State phase = QUESTION\_REMOTE 
                  \Else
                    \State \textsc{VoteToHalt}() \emph{\Comment{All done}}
                  \EndIf
		\EndIf
		\EndProcedure
		
	\end{algorithmic}
\end{algorithm}

\section{Results}
\label{sec:results}

In this section, we present empirical results comparing the vertex and subgraph-centric Triangle Counting algorithms, using Apache Giraph and our own GoFFish platforms, respectively. We leave evaluation of k-Way Clustering and MSF as future work. We run our experiments on $4$ nodes of a commodity cluster, with each node having one AMD Opteron 3380 (8~cores, 2.6~GHz) CPU, 32~GB RAM and 256~GB SSD, inter-connected by Gigabit Ethernet. The nodes run CentOS 7, Giraph v1.2.0, Hadoop v1.2.1 and GoFFish v2.6, on OpenJDK v1.7.0. We use the triangle counting algorithm implemented in Giraph by the Okapi ML library, and implement the proposed subgraph-centric algorithm in GoFFish ourselves. We run these algorithms on three real world graphs from the SNAP repository, after converting each to an undirected graph required by the algorithm~\footnote{Both Giraph and GoFFish represent undirected edges as a pair of directed edges.}. These graphs are listed in Table~\ref{tbl:graphspecs} 

\begin{table}[h]
	\centering
	\caption{Graphs used in Experiments} 
	\begin{tabular}{l|c||rr}
		\hline
		\textbf{Graph Name} & \textbf{Code} &  \textbf{Vertices} & \textbf{Edges}\\
		\hline
		\hline
		California Road Network~$^{1}$ &  CARN & 1,965,206 & 5,533,214\\
		Google Web Graph~$^{2}$ & WEBG & 875,713 & 8,644,102\\
		Patent Citation Network~$^{3}$ & CITP & 3,774,768 & 33,037,895\\
		\hline
	\end{tabular}
$^{1}$~http://snap.stanford.edu/data/roadNet-CA.html \\
$^{2}$~http://snap.stanford.edu/data/web-Google.html \\
$^{3}$~http://snap.stanford.edu/data/cit-Patents.html
	\label{tbl:graphspecs}
\end{table}

\begin{figure} 
\centering
	\includegraphics[width=0.8\columnwidth]{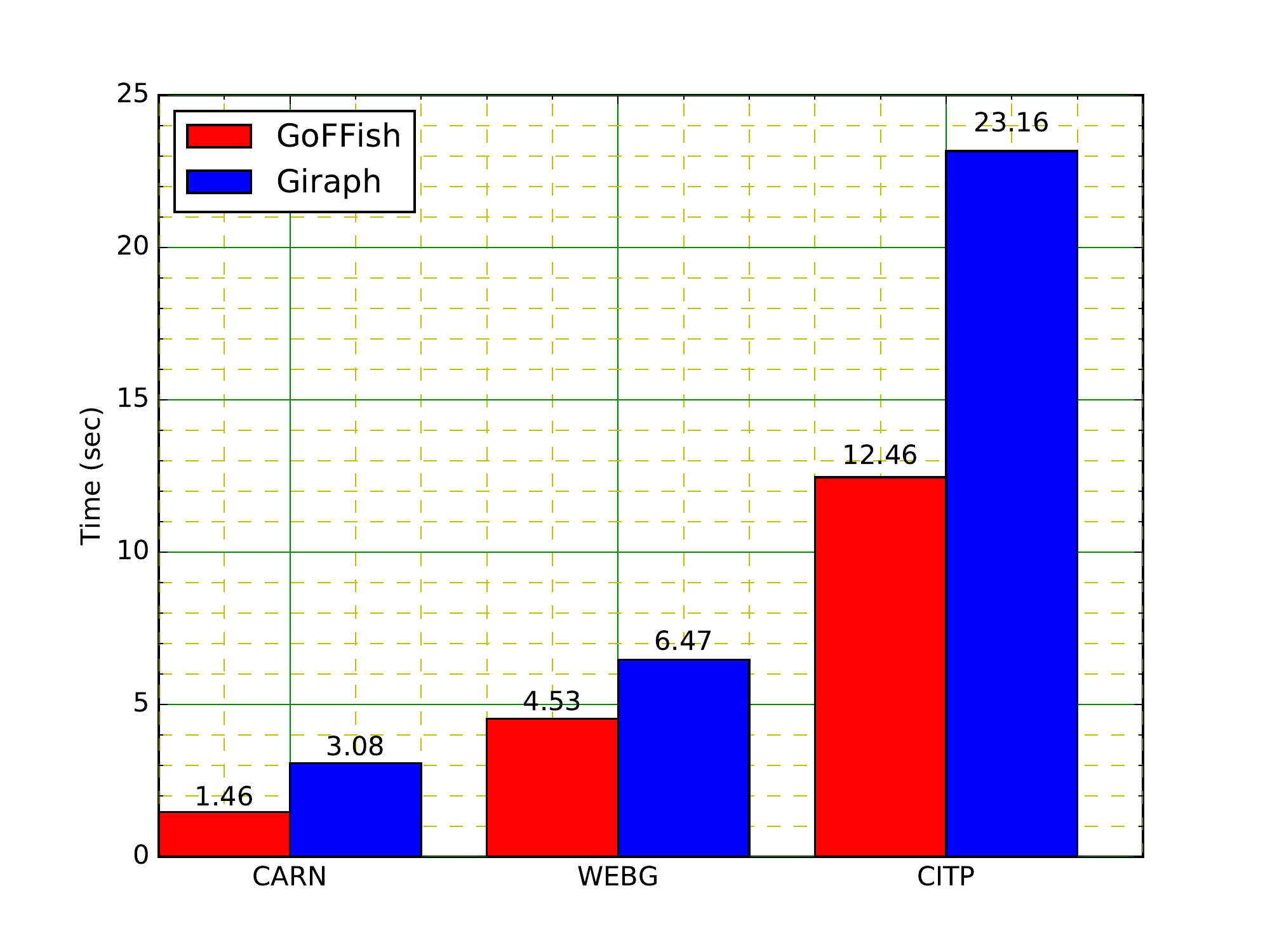}
	\caption{Execution time of triangle counting using GoFFish and Giraph for 3 graphs}
	\label{plot:timings}
\end{figure}

Figure~\ref{plot:timings} shows the time taken by GoFFish and Giraph for the triangle counting algorithm on these three graphs. We see that for both the frameworks, the time taken is a function of the number of edges -- {the communication cost dominates for Giraph, and is a function of the number of edges, and for GoFFish, the computational cost is a function of the local edges while the communication cost depends on the number of remote edges.} So for both platforms, CARN is faster than WEBG even though WEBG has less than half the number of vertices as CARN. CITP having a much larger number of edges than the other two graphs took considerably longer as well, confirming our analysis. 

The execution time for GoFFish is half as that of Giraph for both CARN and CITP, while it is 30\% faster for WEBG. This shows the improvements offered by a subgraph-centric algorithm over a vertex-centric one. We also note that for the CITP graph, GoFFish was able to find all triangles in just two supersteps instead of three since there were no messages generated by the $2^{nd}$ superstep as the test in Line \ref{edge-existence} of Algorithm~\ref{alg:triangle-count} fails. Graphs in GoFFish are pre-partitioned across nodes and we use METIS to perform this partitioning. In the case of CITP, this partitioning was of a good quality such that all the triangles are contained completely within subgraphs (i.e. Types (i) or (ii)), there are no Type (iii) triangles nor are there open triangles that span three subgraphs. 
This gives an added advantage on top of Giraph that always requires 3 supersteps to complete, increasing the coordination overhead.

\section{Conclusion}
\label{sec:conclusion}
In this paper, we designed three algorithms for subgraph-centric frameworks which used their counterpart algorithms from shared-memory and vertex-centric programming models. Our empirical results for triangle counting shows that the subgraph-centric algorithm did perform better than the vertex-centric algorithm for the three graphs we considered. 

As part of future work, we plan to implement and compare the k-way and MSF algorithms with their vertex-centric variants. Further, we propose to identify and classify a broad class of graph algorithms that are well-suited for a subgraph-centric model, and compare them with vertex-centric models using both analytical and empirical methods. This will help map classes of algorithms to the programming model that is best suited to implement it.



\bibliographystyle{IEEEtran}
\footnotesize{
\bibliography{paper}
}

\end{document}